\documentclass[preprint,english,aps,floats,showpacs,nofootinbib]{revtex4}
\usepackage{pslatex}
\usepackage[T1]{fontenc}
\usepackage[latin1]{inputenc}
\usepackage{graphicx}
\usepackage{epsfig}

\usepackage{calc}
\usepackage{ifthen}

{\newcommand{\lsim}{\mbox{\raisebox{-.6ex}{~$\stackrel{<}{\sim}$~}}}
{ 
{
\newcommand{\be}{\begin{equation}}
\newcommand{\ee}{\end{equation}}
\newcommand{\bea}{\begin{eqnarray}}
\newcommand{\eea}{\end{eqnarray}}

\newcommand{\nc}{\newcommand}
\nc{\eqj}[1]{\mbox{Eq.~(\ref{#1})}}
\nc{\bej}[1]{\begin{equation} \mbox{$\label{#1}$}}
\nc{\eej}{\vspace{0.1cm}\end{equation}}

\newcommand{\bean}{\begin{eqnarray*}}
\newcommand{\eean}{\end{eqnarray*}}

\def\GeV{{\rm \ GeV}}
\def\MeV{{\rm \ MeV}}

\def\TeV{{\rm \ TeV}}

\def\lae{\;^{<}_{\sim} \;} \def\gae{\; ^{>}_{\sim} \;}

\begin{document}
\title{Cosmic Ray Anomalies and Dark Matter Annihilation to Muons via a Higgs Portal Hidden Sector}
\author{Kazunori Kohri}
\email{k.kohri@lancaster.ac.uk}
\author{John McDonald}
\email{j.mcdonald@lancaster.ac.uk}
\author{Narendra Sahu}
\email{n.sahu@lancaster.ac.uk}
\affiliation{Cosmology and Astroparticle Physics Group, University of 
Lancaster, Lancaster LA1 4YB, UK}
\begin{abstract}

    Annihilating dark matter (DM) models based on a scalar hidden sector with Higgs 
portal-like couplings to the Standard Model are considered as a possible explanation 
for recently observed cosmic ray excesses. Two versions of the model are studied, one 
with non-thermal DM as the origin of the boost factor and one with Sommerfeld 
enhancement. In the case of non-thermal DM, four hidden sector scalars which transform 
under a $U(1)_{X}$ symmetry are added. The heaviest scalars decouple and later decay to DM 
scalars, so providing the boost factor necessary to explain the present DM annihilation rate. 
The mass of the annihilating scalars is limited to $\lae 600$ GeV for the model to remain 
perturbative.
$U(1)_{X}$ breaking to $Z_2$ at the electroweak transition mixes light O(100) MeV  
hidden sector scalars with the Higgs. The DM scalars annihilate to these light scalars, which 
subsequently decay to two $\mu^{+}\mu^{-}$ pairs via Higgs mixing, so generating a positron excess 
without antiprotons. Decay to $\mu^{+}\mu^{-}$ rather than $e^{+}e^{-}$ is necessary to ensure a 
fast enough light scalar decay rate to evade light scalar domination at nucleosynthesis. In the 
version with Sommerfeld enhancement only three new scalars are necessary. TeV scale DM masses 
can be accomodated, allowing both the higher energy electron plus positron excess and the 
lower energy PAMELA positron excess to be explained. DM annihilates to 2$\mu^{+}\mu^{-}$ 
pairs as in the non-thermal model. This annihilation mode may be favoured by recent observations of 
the electron plus positron excess by FERMI and HESS.

\end{abstract}
\maketitle

\section{Introduction}

Recent results from the satellite experiment PAMELA~\cite{pamela_positron} indicate an excess 
of positrons at 10-100 GeV as compared with the expected galactic background, confirming the 
earlier results from HEAT~\cite{heat_data} and AMS~\cite{ams_data}. Surprisingly, PAMELA did 
not find any antiproton excess below 100 GeV as compared with the galactic background~\cite{
pamela_proton}. Evidence was obtained from the balloon experiments ATIC~\cite{atic_result} and
PPB-BETS~\cite{ppb_result} of an excess electron plus positron flux as compared with the 
galactic background in the energy range 100-800 GeV. These results have recently been reconsidered 
by FERMI \cite{fermi} and HESS \cite{hess}, which do not confirm the large excess and
spectral features observed by ATIC and PPB-BETS. However, HESS does not rule
out the possibility of an electron plus positron excess, although there is no
indication of structure in the electron plus positron spectrum \cite{hess}, while
FERMI observes a flattening of the electron spectrum relative to that
predicted by a conventional diffusive model for the background, which
suggests new physics, although again no prominent spectral features
are observed \cite{fermi}. Therefore an electron plus positron excess remains a
possibility. These results raise the exciting prospect that the
positron and the electron plus positron excesses could be attributed to annihilation of
dark matter (DM) particles\footnote{ Nearby astrophysical sources~\cite{neutralino_clump,
pulsar,Astro3} and decaying DM~\cite{DecayFermi,thaler} are also possible 
explanations.}. If DM annihilation is the explanation for the positron excess at lower 
energies and the possible electron plus positron excess at higher energies, then the 
annihilation rate of DM at present should be larger than that expected from the canonical 
thermal relic annihilation cross-section in the case of a smooth distribution of DM in 
the galaxy ($\approx 3 \times 10^{-26}$ ${\rm cm}^3/{\rm s}$)~\cite{cireli:08}. This is 
the boost factor\footnote{Different authors define the boost factor in different ways, with 
some reserving this term for the astrophysical boost due to clumpy DM. We will
use it to refer to the total enhancement of the DM annihilation rate.}. The origin of the 
boost factor could be astrophysical, because of the merger of sub-structures, or entirely
from particle physics, or a combination of the two.

     A popular method to achieve the boost factor is Sommerfeld enhancement of the DM 
annihilation cross-section ~\cite{cireli:08,se,se2,se3}. This typically requires the 
introduction of new light bosons of mass $M_{B} \sim \alpha M_{DM}$ in order to mediate 
a force between the DM particles, where $M_{DM}$ is the DM particle mass and $\alpha$ is 
the interaction's fine-structure constant. (An exception is discussed in \cite{se}, where 
the enhancement is mediated by electroweak interactions.) An alternative approach is to 
use non-thermal production of DM to accomodate a large annihilation cross-section, usually 
via heavy particle decay\footnote{We note that in SUSY models a natural alternative is Q-ball 
decay \cite{qb1,qb2}.} \cite{fairbairn:2008}. In the following we will consider both 
possibilities\footnote{Another possibility, annihilation close to a pole, has been considered in \cite{bw1,bw2}.}.

In addition, in order to produce positrons without a sizable amount
of antiprotons, a mechanism to allow DM to annihilate primarily to leptons is required. One 
approach is to introduce new `leptophilic' couplings of the DM particles to leptons
\cite{leptonic}. In this paper we will instead consider a DM sector which interacts with 
the SM via generic (non-leptophilic) couplings.  Our goal is to determine whether such 
non-leptophilic models can account for the observed cosmic ray excesses and to obtain the 
necessary conditions on their masses, couplings and field content. 
Our analysis of the ingredients required to construct successful non-leptophilic models may then guide the construction
of more complete models which can explain the necessary features.

    Our models are based on a SM singlet scalar sector interacting with the SM via Higgs 
portal-like interactions. Adding a scalar DM particle $S$ is a particularly simple way to 
extend the SM to account for DM~\cite{sdm1,sdm2}. The scalar is typically stabilised by 
either a discrete $Z_{2}$ or $U(1)$ symmetry. It interacts with the SM sector via the 
coupling $S^{\dagger}S H^{\dagger}H$ (which has come to be known as the `Higgs portal' 
\cite{portal}), which is the only renormalizable coupling of the $S$ to SM particles. 
Several DM models based on this type of coupling have been proposed \cite{sdm}. However, 
this coupling alone cannot account for DM annihilation primarily to leptons, nor can it 
account for the boost factor. Here we extend the symmetry of the DM particle to a sector 
of SM singlet scalar fields.  The DM sector is composed of the DM scalar $S$ and 
additional scalar fields $\chi_i$, all of which carry non-trivial charges under a 
symmetry $U(1)_X$. In the model with non-thermal DM, three scalars $\chi_{i}, i=1,2,3$,  
are introduced. The heaviest scalar $\chi_1$ populates the number density of DM, so providing the boost factor, 
while the lightest scalar $\chi_3$ ensures an annihilation channel of DM to two $\mu^+ \mu^-$ 
pairs. The role of $\chi_2$ is to mix $\chi_{3}$ with the Higgs via its vacuum expectation 
value (VEV), inducing its decay to leptonic final states.
In order to ensure that $\chi_3$ decays before dominating at nucleosynthesis, $U(1)_{X}$ 
must be broken to a discrete $Z_{2}$ symmetry which maintains the stability of $S$. This 
occurs spontaneously at the electroweak phase transition, when $\chi_2$ acquires a VEV. 
$U(1)_{X}$ breaking also mixes the lightest scalar $\chi_{3}$ with the Higgs boson, providing 
a mechanism for leptonic annihilation of $S$ to 2 $\mu^{+}\mu^{-}$ pairs\footnote{The Higgs 
mixing mechanism was first described in \cite{se2}.}. In this version of the model the mass of $S$
is constrained by perturbativity to be less than approximately 600 GeV. In the version of the model with Sommerfeld enhancement, the light scalar $\chi_{3}$ which mixes with the Higgs is also used to mediate the enhancement. 
This model requires only two scalars, $\chi_{2}$ and $\chi_{3}$, and can accomodate a larger range of $S$ mass.  
DM again annihilates to 2 
$\mu^{+}\mu^{-}$ pairs via Higgs mixing as in the model with non-thermal DM. Annihilation 
to muons appears to be favoured by recent data from FERMI and HESS~\cite{na1,na2}, with 
annihilation to 4$\mu$ being favoured by the analysis of~\cite{na2}.  

    Our paper is organised as follows. In Section II we present and discuss a Higgs 
portal model with non-thermal production of DM as the source of the boost factor. 
In Section III we discuss a version of the model with Sommerfeld enhancement in place of 
non-thermal production. In Section IV we present our conclusions.

\section{A Higgs Portal Model with Non-Thermal Dark Matter}

\subsection{Overview}

   We extend the SM by adding a dark sector composed of a singlet DM scalar $S$ 
of mass $M_S$ and three additional scalar fields $\chi_i$, $i = 1,2,3$ with masses 
$m_i$, $i=1,2,3$, such that $m_1 \gg m_2 \gg m_3$. We impose a symmetry $U(1)_X$, under 
which the fields $\chi_i$ carry a charge $+$1 and $S$ carries a charge $+$3/2. In 
order to avoid a Goldstone boson from $U(1)_{X}$ breaking we will consider this to 
be a gauge symmetry. The gauge interaction will not have any significant effect on the cosmological 
evolution of the model, only contributing to the already rapid annihilation and scattering between the hidden 
sector scalars.  
The SM fields are neutral under $U(1)_X$. The dark sector 
fields interact with the SM via Higgs portal-type couplings to the Higgs bilinear 
$H^{\dagger}H$. $U(1)_X$ will be broken at the electroweak (EW) phase transition to a 
surviving $Z_2$ symmetry under which $S$ is odd while rest of the fields, including the 
SM fields, are even. Since $S$ is odd under the surviving $Z_2$ symmetry, it is stable 
and a candidate for DM. 

Our model is based on generic couplings of the gauge singlet scalars to 
$H^{\dagger}H$. The renormalizable couplings of the scalar sector of the Lagrangian 
are given by
\bea
{\mathcal L}  & \supseteq &  m_{i}^2 \chi_i^\dagger \chi_i + M_S^2 S^\dagger S + 
m_H^2 H^\dagger H \nonumber\\
& + & \lambda_S (S^\dagger S)^2  + \lambda_H (H^\dagger H)^2 + \gamma S^\dagger S 
H^\dagger H   \nonumber\\
& + & \eta_{ijkl} \chi_i^\dagger \chi_j \chi_k^\dagger \chi_l + (\eta_S)_{ij} \chi_i^\dagger 
\chi_j S^\dagger S +  (\eta_H)_{ij} \chi_i^\dagger \chi_j H^\dagger H
~,
\label{lagrangian}
\eea
where $H$ is the SM Higgs and $i = 1,2,3$. We assume that all couplings are real and 
that all particles in the dark sector have positive masses squared ($m_i^2, M_S^2 > 0$). As 
usual $m_H^2 < 0$ so that $H$ acquires a vacuum expectation value (VEV). If $(\eta_{H})_{22} < 0$ 
and $m_{2}^{2}$ is sufficiently small then $\chi_{2}$ gains a $U(1)_{X}$-breaking VEV when the 
electroweak (EW) phase transition occurs.  

   The cosmological evolution of the model can be summarized as follows. $\chi_{1}$ is 
assumed to have the largest mass in the hidden sector. $S$ has a mass $\gae$ 100 GeV if 
its annihilation is to  account for the PAMELA observations and so freezes-out at a 
temperature $\gae 5$ GeV. A key requirement of the model is that $\chi_{1}$ decay occurs 
sufficiently long after the $S$ density has frozen out of thermal equilibrium that it 
can boost the $S$ relic density. $\chi_{1}$ 
can decay to $\chi_{i}^{\dagger} \chi_{j}\chi_{k}$, $\chi_{i} S^{\dagger}S$ or $\chi_{i} 
H^{\dagger}H$ ($i,j,k\neq 1$). After the EW phase transition, when $\chi_2$ acquires a VEV, 
$\chi_1$ also decays through the two body processes: $\chi_1 \rightarrow 
\chi_j^\dagger \chi_k$, with $j,k \neq 1$ and $\chi_1\rightarrow S^\dagger S, h h, \chi_j h$ 
with $j \neq 1$, where $h$ is the physical Higgs scalar. We will see that late decay of 
$\chi_1$ requires these couplings to be very small, of order $10^{-10}$.  
Due to $U(1)_X$-breaking, $\chi_{3}$ mixes with the physical Higgs scalar $h$ and decays to SM 
fermions via the Yukawa couplings. If the $\chi_{3}$ mass is in the range $2 m_{\mu}$ to 
$2 m_{\pi^{o}}$ (212-270 MeV) then it decays predominantly to $\mu^{+} \mu^{-}$ pairs. For 
larger $\chi_3$ mass, $\chi_3$ decay to $\pi^{o}$ pairs produces photons while $\chi_3$ decay 
to nucleon-antinucleon pairs produces antiprotons. Whether the photon flux from pion decay 
is excluded depends on the nature of the DM halo, with cuspy NFW halos excluded 
but cored isothermal halos still likely to be consistent with present bounds\footnote{A 
possible problem with photons from pion decay was previously noted in the context of an axion 
decay model for cosmic ray anomalies~\cite{thaler}.}~\cite{Bertone:2008xr,Kawasaki:2009nr}. 
In the following we will consider the $\chi_3$ mass to be in the range $2 m_{\mu}$ to 
$2 m_{\pi^{o}}$, although the upper bound may be increased to 2$m_{proton}$ if the photon 
flux from pion decay is within observational limits. This range of $\chi_{3}$ mass requires 
that the couplings of $\chi_{3}$ to $H$ and $\chi_{2}$ are less than $10^{-6}$. The lifetime 
for $\chi_{3}$ decay to $\mu^{+}\mu^{-}$ is short compared with the time at which 
nucleosynthesis begins, so the relic $\chi_{3}$ density, which would otherwise dominate at 
nucleosynthesis, safely decays away. This would not be true for decay to $e^{+}e^{-}$, which 
would apply if the $\chi_{3}$ mass was less than $2 m_{\mu}$. The present $S$ density annihilates 
primarily to $\chi_{3}$ pairs which promptly decay to muons. The subsequent decay  $\mu^{+} 
\rightarrow e^{+} + \nu_{e} + \overline{\nu}_{\mu}$ then accounts for the positron flux without 
any antiproton flux.

\subsection{Electroweak Phase Transition and Spontaneous Breaking of $U(1)_X$}

  After the EW phase transition $H$ develops a VEV, which triggers a VEV for $\chi_2$.
The VEV of $H$ and $\chi_2$ also induce a VEV for $\chi_1$ through the
couplings $\eta_{1222}$ and $(\eta_H)_{12}$ 
\be
 \langle \chi_1 \rangle \approx - \frac{\eta_{1222}\; u^3 + (\eta_H)_{12} \;u 
v^2}{m_1^2} 
\label{vev_u1}
\ee
where $\langle \chi_2 \rangle = u$ and 
$\langle H \rangle=v$. As we will show, $\eta_{1222}$ and 
$(\eta_H)_{12}$ are required to be no larger than ${\rm O}(10^{-10})$ in order to ensure the late 
decay of $\chi_1$. Therefore with $u \sim v \sim 100$ GeV and $m_1 \sim {\rm O}(1)$ TeV we find that $\langle \chi_1 \rangle$ is negligibly small, ${\rm O} (100)$ eV. Similarly, $\chi_{3}$ also gains a VEV   
\be
\langle \chi_3 \rangle \approx - \frac{\eta_{2333}\; u^3 + (\eta_H)_{23} \;u 
v^2}{m_3^2} 
\label{vev_u3}
~.\ee
We will show later that the mass of the lightest mass eigenstate $\chi_{3}^{'}$ 
must be less than O(1) GeV in order to ensure that it will decay primarily to 
leptons. 
This is most easily understood if all the terms in the mass matrix are less than $1 \GeV$, which in turn
requires that all the couplings of $\chi_{3}$ to $\chi_{2}$ and $H$ are less than O$(10^{-6})$. (Larger entries in the mass matrix are possible but would require sufficient cancellation between the contributions to the lightest mass eigenvalue.)
Since $m_{3}$ is also no larger than O(1) GeV, this means that $\langle \chi_{3} \rangle \lae u,\; v$. 
Although a value for $<\chi_{3}>$ which is comparable to $\langle H \rangle$ and $\langle \chi_2 \rangle$ is possible, this will not qualitatively alter our results from the case where $\langle \chi_{3} \rangle \ll u,\;v$, since it will only alter the admixtures of $\chi_2$ and $H$ in the lightest mass eigenstate by O(1) factors. Therefore, to simplify the analysis we will set $<\chi_{3}> = 0$ in the following and consider
\bej{e1}
\langle \chi_1 \rangle = 0,\, \;\; \langle S \rangle= 0,\,\;\; \langle H \rangle =v,\, \;\;
\langle \chi_2 \rangle= u\, \;\;  \langle \chi_3 \rangle=0\;,
\eej
with $u$ and $v$ obtained by minimizing the scalar potential
\bej{e2}
V=m_H^2 v^2 + m_2^2 u^2 +\eta_{2222} u^4 + \lambda_H v^4 + (\eta_{H})_{22} u^2 v^2 \,.
\eej
We assume that $(\eta_{H})_{22}$ and $m_{H}^2$ are negative with all other terms positive. Vacuum stability requires that $(\eta_{H})_{22} > -2 \sqrt{\eta_{2222} \lambda_{H}}$.  
Minimizing \eqj{e2} gives
\bej{e3}
u=\sqrt{\frac{( 2 \lambda_H m_2^2 - (\eta_{H})_{22} m_H^2)}{
((\eta_{H})_{22}^2 -4 \eta_{2222} \lambda_H)}}
\eej
and
\bej{e4}
v=\sqrt{\frac{(2\eta_{2222} m_H^2 -(\eta_{H})_{22} m_2^2)}{((\eta_{H})_{22}^2 - 4\eta_{2222} \lambda_H)}}
~.\eej
In the following we will assume that $\eta_{2222}$, $|(\eta_{H})_{22}|$ are $\sim 0.1$ and that $m_{2} \lae 100 \GeV$, therefore $u \sim v$.   

       The $\chi_{2}$ expectation value breaks the $U(1)_X$ symmetry to a discrete symmetry, under 
which the scalars transform as $\phi \rightarrow e^{i 2 \pi Q} \phi$, where $Q$ is the $U(1)_X$ 
charge of $\phi$. Thus $\chi_{i}$ ($Q = 1$) transforms as $\chi_i \rightarrow \chi_i$, while $S$ 
($Q = 3/2$) transforms as $S \rightarrow -S$. Therefore $S$ is stable due to the residual discrete 
symmetry.

   Once $\chi_{2}$ gains a $U(1)_{X}$-breaking expectation value, the Higgs $h$ (defined by $Re(H^{0}) \equiv (h + v)/\sqrt{2}$) mixes with the real parts of $\chi_1$, $\chi_{2}$ and $\chi_{3}$. The $\chi_{2}$ expectation value is assumed to be $\sim v$, so $\chi_{2}-h$ mixing will be large, which should have significant consequences for Higgs phenomenology.  
$\chi_{3}$-$h$ mixing provides the mechanism for DM to annihilate primarily to lepton final states. For simplicity we will consider only the mixing of the Higgs with 
$\chi_{3}$, which is responsible for the important physics. In the basis spanned by $\sqrt{2} {\rm Re} \chi_3$ and $h$, 
the effective mass squared matrix is given by
\bej{e5}
{\cal M}^2 = \pmatrix {m_3^2 + 6 \eta_{2233} u^2 + (\eta_{H})_{33} v^2 
& (\eta_H)_{23} u v \cr
(\eta_H)_{23} u v & 2 \lambda_H v^2 + (\eta_{H})_{22} u^2 }
~.\eej
In this we have assumed that $\eta_{ijkl}$ is independent of the order of $i,j,k,l$. 
Diagonalising this gives mass eigenstates $\chi_{3}^{'}$ and $h^{'}$
with
\bej{e6} M_{\chi_{3}^{'}}^{2} \approx m_3^2 + 6 \eta_{2233} u^2 + (\eta_{H})_{33} v^2  
- \frac{((\eta_H)_{23} u v)^{2}}{M_{h^{'}}^{2}} ~\eej
and 
\bej{e7} M_{h^{'}}^2 \approx  2 \lambda_H v^2 + (\eta_{H})_{22} u^2  ~,\eej
where we assume $\chi_{3} - h$ mixing is small. The $\chi_{3}-h$ mixing angle is 
\bej{e8} \beta \approx \frac{(\eta_{H})_{23} uv}{M_{h^{'}}^2}   ~.\eej
In order to have DM annihilation to muon final states we require that 
$M_{\chi_{3}^{'}}$ is in the range 212-270 MeV. Therefore we require that $m_3 \lae 300 \MeV$, 
$\eta_{2233} \lae 10^{-6}$, $(\eta_{H})_{33} \lae 10^{-6}$ and $(\eta_{H})_{23} \lae 10^{-6}$, 
assuming that $u \sim v \sim M_{h^{'}} \sim 100 \GeV$. Although for simplicity we considered 
only the mixing between $\chi_3$ and $h$, in general $\chi_3$ will mix with $\chi_2$ and 
$\chi_1$ in addition to $\chi_3$. The corresponding couplings will also be constrained by 
the requirement that the light eigenstate mass is $\sim {\rm O} (100)$ MeV,
therefore the additional mixings will not change the model qualitatively. This illustrates an 
important feature of generic Higgs portal models for cosmic ray excesses: some couplings must 
be strongly suppressed. We will show that suppressed couplings are also necessary to produce 
the boost factor via non-thermal production of DM.

\subsection{Non-Thermal Production of $S$ Dark Matter}

    The $S$ density is due to out-of-equilibrium decay of $\chi_1$. This must occur at a 
sufficiently low temperature that the $S$ scalars can have a boosted annihilation cross-section 
without annihilating away after being produced by $\chi_{1}$ decay. An initial density produced 
by $\chi_{1}$ decay at temperature $T_{decay}$ will annihilate down to a density at $T_{decay}$ 
given by 
\bej{a1}     n_S(T_{decay}) = \frac{H(T_{decay})}{(\sigma |v_{\rm rel}|)_S}     ~,\eej
where $n_S$ is the number density and $(\sigma |v_{\rm rel}|)_S$ is the annihilation cross-section 
times relative velocity (which is $T$ independent for the case of annihilating scalars). 
\eqj{a1} is true if the initial $S$ number density from $\chi_{1}$ decay is larger than 
$n_S(T_{decay})$. Since $n_{S} \propto g(T) T^3$ while $H \propto g(T)^{1/2} T^2$, where $g(T)$ 
is the effective number of relativistic degrees of freedom, \eqj{a1} implies that the 
present ratio of the $S$ number density from $\chi_{1}$ decay to that from thermal 
freeze-out (which is given by 
\eqj{a1} with $T_{decay}$ replaced by the $S$ freeze-out temperature $T_{S}$) is  
\be \frac{n_{S\;decay}}{n_{S\;th}}  =  \left( \frac{g(T_{S})}{g(T_{decay})} \right)^{1/2}  
\frac{T_{S}}{T_{decay}}  ~.\ee 
Since the annihilation cross-section is enhanced by a factor $B$ over that which accounts for 
observed DM via a thermal relic density, it follows that the thermal relic $S$ density is smaller 
than that observed by a factor $B$. So in order to account for the observed DM via $\chi_{1}$ 
decay we must require that 
$n_{S\;decay} \approx B \times n_{S\;th}$. Therefore
\bej{a1b} T_{decay} = \left( \frac{g(T_{S})}{g(T_{decay})} \right)^{1/2} \frac{T_{S}}{B} ~.\eej
In this we have neglected the logarithmic dependence of the freeze-out temperature on the 
annihilation cross-section and therefore treated $T_{S}$ as a constant. $T_{S}$ is related to the 
$S$ mass by $T_{S} = M_{S}/z_{S}$ with $z_{S} \approx 20$ \cite{leeweinberg}.  
For example, if $m_{S} \sim 400 \GeV$ then $T_{S} \sim 20 \GeV$. Since the observed positron and 
electron excess requires that $B \sim 10^3$, we would then require that $T_{decay} \sim 20 \MeV$.

  This very low decay temperature is difficult to achieve via particle decay, as it implies a very long-lived particle with lifetime $\tau \sim H^{-1} \sim 10^{-3} s$.  $\chi_{1}$ can decay to $S$ pairs via the three-body decays  $\chi_1 \rightarrow \chi_2 S^\dagger S$ and $ \chi^{'}_{3} S^\dagger S$. 
The decay rate is given by 
\be
\Gamma_{\chi_1} \approx \frac{\lambda^2}{128 \pi^3} M_{\chi^{'}_1}
\label{threebody-decay}
~,\ee
where $\lambda^2=(\eta_S)_{12}^2+(\eta_S)_{13}^2$. In order to have a late enough $\chi_{1}$ decay we require that $\Gamma_{\chi_{1}} < H$ at 
$T = {\rm O}(10) \MeV$, which implies that $\lambda \lae 10^{-10}$.   
$\chi_{1}$ can also decay to $S$ pairs via the two-body decay $\chi_1 \rightarrow S^\dagger S$ once $\chi_2$ gains a VEV. The decay rate is given by  
\be
\Gamma_{\chi_1} \approx \frac{(\eta_S)_{12}^2}{16\pi} \frac{u^2}{M_{\chi_1'}}  \label{twobody-decay}
~.\ee
With $u \sim 100 \GeV$ and $M_{\chi_1'} \sim 1 \TeV$, this decay rate as a function of $(\eta_S)_{12}$ is comparable 
to Eq.(\ref{threebody-decay}) as a function of $\lambda$. Therefore we also require that $(\eta_S)_{12} \lae 10^{-10}$.   
(Similar constraints apply to other $\chi_1$ two-body decay modes.) 

    The above assumes that the initial $S$ density from $\chi_{1}$ decay is larger than that 
given in \eqj{a1}. This requires that $n_{\chi_{1}}(T_{decay}) > n_{S}(T_{decay})/2$, since 
each $\chi_{1}$ decay produces 2 $S$. $n_{\chi_{1}}(T_{decay})$ is given by
\bej{a2} n_{\chi_{1}}(T_{decay}) = \frac{g(T_{decay})}{
g(T_{\chi_{1}})}
\frac{T_{decay}^3}{ T_{\chi_{1}}^3} \frac{H(T_{\chi_{1}})}{(\sigma |v_{\rm rel}|)_{\chi_{1}}}     ~,\eej
where $T_{\chi_{1}}$ is the $\chi_{1}$ freeze-out temperature.  
The condition $n_{\chi_{1}}(T_{decay}) > n_{S}(T_{decay})/2$ then implies that
\bej{a3} \frac{g(T_{decay})^{1/2}}{g(T_{\chi_{1}})^{1/2}} 
\frac{T_{decay}}{T_{\chi_{1}}} > \frac{1}{2} \frac{(\sigma |v_{\rm rel}|)_{\chi_{1}}}{ (\sigma |v_{\rm rel}|)_{S}}   ~.\ee
This translates into an upper bound on $B$, 
\bej{a5} B < B_{0} \equiv \frac{ 2(\sigma |v_{\rm rel}|)_{S}}{(\sigma |v_{\rm rel}|)_{\chi_{1}}} \frac{z_{\chi_{1}}}{z_{S}} 
\frac{M_{S}}{M_{\chi_{1}}} \frac{g(T_{S})^{1/2}}{g(T_{\chi_{1}})^{1/2}}     ~.\eej
Therefore if $B < B_0 \approx (\sigma |v_{\rm rel}|)_{S}/(\sigma |v_{\rm rel}|)_{\chi_{1}}
$ (assuming that $M_{\chi_{1}} \sim M_{S}$ and $z_{\chi_{1}} \sim z_{S}$) then the required $\chi_1$ decay temperature will be given by \eqj{a1b}. The cross-section times relative velocity for non-relativistic $\chi_{1}$ 
pair annihilation to $S$ and $H$ is given by    
\be (\sigma |v_{\rm rel}|)_{\chi_1} = \frac{1}{32 \pi M_{\chi_1}^2}\left[ (\eta_{S})_{11}^2 + (\eta_{H})_{11}^2  \right]\;
\label{sigma_chi1}
~.\ee
The annihilation cross-section times relative velocity for non-relativistic S is given by  
\be
(\sigma|v_{\rm rel}|)_S = \frac{1}{32 \pi M_{S}^2}\left[ (\eta_{S})_{ij}^2 + \gamma^2 \right]\;\;\;,\;  i = 2,3\;\;
\label{sigma-s2}
~.\ee
With $B \sim 10^2-10^3$, \eqj{a5} is therefore satisfied if $(\eta_{H})_{11}, 
(\eta_{S})_{11} \lae 10^{-2}$, assuming that $\gamma \sim 0.1$. 
If this is not satisfied and $B > B_{0}$, then the boost factor is given by $B_{0}$ rather 
than $B$. In this case the $S$ density comes directly from $\chi_{1}$ decay without subsequent 
annihilations. An even lower $\chi_{1}$ decay temperature and smaller $(\eta_S)_{12}$ would then 
be necessary in order to account for the observed DM density.

\subsection{Dark Matter Annihilation Rate} 

The dominant $S$ annihilation mode is assumed to be to $\chi_{3}^{'}$ pairs. In this case the 
annihilation cross-section times relative velocity is given by 
\be
(\sigma|v_{\rm rel}|)_S = \frac{(\eta_{S})_{33}^2}{32 \pi M_{S}^2}
\label{sigma-s}
~.\ee
In order to account for the cosmic ray excesses, the annihilation cross-section times 
relative velocity necessary to account for thermal relic DM, $(\sigma |v_{rel}|) 
\approx 3 \times 10^{-26} {\rm cm^3/s} \equiv 2.6 \times 10^{-9} \GeV^{-2}$, must be boosted 
by $B \sim 10^2 - 10^3$ for DM masses in the range O(100)GeV - O(1)TeV. This requires 
that
\be (\eta_{S})_{33} \approx (5-16) \times \left(\frac{M_{S}}{1 \TeV}\right)~.
\ee 
Therefore for the theory to remain perturbative ($(\eta_{S})_{33} \lae 3$) we require that 
$M_{S} \lae 600 \; (190) \GeV$ for $B = 10^2\;(10^3)$. Thus while the model can account for 
the positron excess in the range 1-100 GeV observed by PAMELA, an electron plus positron excess 
at energies up to O(1)TeV cannot be explained if the model is to remain perturbative, in which 
case an alternative explanation for the electron plus positron excess is required, most likely 
astrophysical. This conclusion is likely to apply rather generally to models which do not have 
Sommerfeld enhancement of the annihilation cross-section. (However, this does not exclude the 
possibility of a large annihilation cross-section due to strong coupling between $S$ and 
$\chi_3$, which is only constrained by unitarity~\cite{unitarity}.)

\subsection{Leptonic Final States via Dark Matter Annihilation to $\chi^{'}_3$}

      In order to account for the positron excess without an accompanying antiproton flux, the 
$S$ annihilations at present should proceed primarily through leptonic decay channels. In our 
model this is achieved through a mixture of $\chi_{3}$-Higgs mixing and kinematics. $U(1)_{X}$ 
breaking due to $<\chi_{2}>$ causes the real part of $\chi_3$ to mix with $h$. If the dominant 
$S$ annihilation mode is $S^{\dagger}S \rightarrow \chi_{3}^{'\;\dagger}\chi^{'}_{3}$ and if 
the $\chi^{'}_{3}$ mass is in the range  $2 m_{\mu} < M_{\chi^{'}_{3}} < 2 m_{\pi^{o}}$, then 
the mixing of $\chi_{3}$ with $h$ leads to the decay $\chi^{'}_{3} \rightarrow \mu^{+} \mu^{-}$ 
via the muon Yukawa coupling. This is illustrated in Figure 1, treating the mixing as a mass 
insertion. Thus $S$ annihilation will produce a 4$\mu$ final state via the process shown in 
Figure 2.  

    The decay rate for $\chi^{'}_3 \rightarrow \mu^+ \mu^-$ is given by
\be
\Gamma_{\chi^{'}_3} = \frac{\beta^2 Y_{\mu}^2}{8 \pi}M_{\chi^{'}_{3}} 
\equiv \frac{(\eta_H)_{23}^2 Y_\mu^2}{8 \pi}  \left( \frac{u v}{M_h^2} \right)^2 M_{\chi^{'}_3} 
\,,
\ee
where $Y_\mu$ is the Yukawa coupling of the SM Higgs to $\mu^+ \mu^-$. This gives for the lifetime of
$\chi^{'}_3$
\bea
\tau_{\chi^{'}_3}\equiv \frac{1}{\Gamma_{\chi^{'}_3}} &=&  4 \times 10^{-4} 
\left(\frac{10^{-6}}{(\eta_H)_{23}} \right)^2
\left( \frac{6.07 \times 10^{-4}} {Y_\mu}\right)^2 \nonumber\\
&& \times \left( \frac{M_h}{150 {\rm GeV}} \right)^4
\left( \frac{200 {\rm MeV}}{M_{\chi^{'}_3}} \right) {\rm s}\;,
\eea
where we have used $u=100$ GeV and $v=174$ GeV. The short lifetime of $\chi^{'}_3$ ensures 
that the thermally produced $\chi^{'}_3$ will decay well before the onset of nucleosynthesis 
at O(1)s. This is essential, as the $\chi_{3}^{'}$ will freeze-out while relativistic (since 
there are no annihilation channels for $\chi^{'}_{3}$ once $T \lae M_{\chi_2}, M_{h^{'}}$) and 
so they will dominate the energy density at nucleosynthesis. The decay rate to $e^{+}e^{-}$ is 
suppressed by $Y_{e}^2/Y_{\mu}^2 \sim  10^{-5}$, leading to a lifetime $\sim 10$ s. Thus for 
$\chi_{3}^{'}$ to decay before nucleosynthesis, {\it decay to muon pairs must be kinematically 
allowed}\footnote{We note that the mechanism described in \cite{se2}, which is based on a single scalar field $\phi$ and which gives a decay rate equivalent to that here but with $<\chi_{2}>$ replaced by $<\chi_{3}> \sim 200 \MeV$ (where $\chi_{3}$ is the equivalent of $\phi$ in 
\cite{se2}), results in a lifetime which is too long and so $\phi$ domination at nucleosynthesis.}. 
\begin{figure}[h]
\begin{center}
\epsfig{file=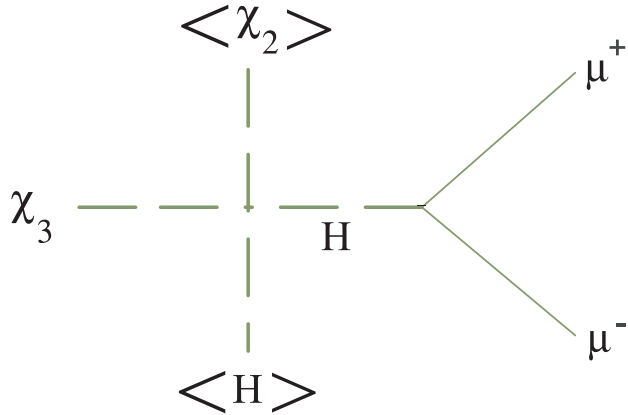, width=0.5\textwidth}
\caption{$\chi_{3}^{'}$ decay into $\mu^{+} \mu^{-}$ via $\chi_3$-Higgs mixing and the muon Yukawa coupling.}
\label{chi3}
\end{center}
\end{figure}
\begin{figure}[h]
\begin{center}
\epsfig{file=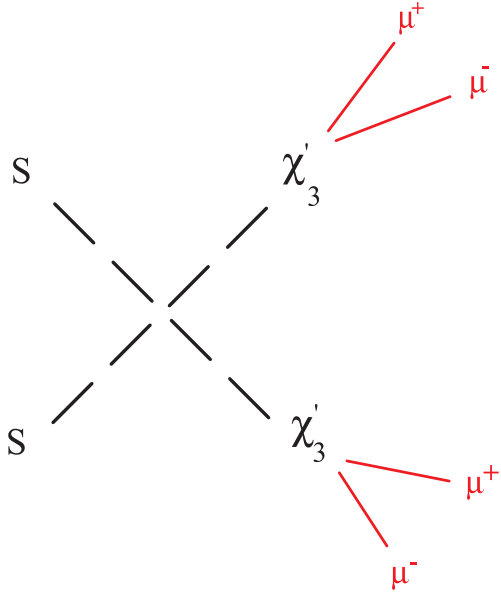, width=0.4\textwidth}
\caption{The primary $S$ annihilation process to 4$\mu$ via $\chi_{3}^{'}$ decay.
}
\label{Sann}
\end{center}
\end{figure}

      The previous discussion applies to the decay of the real part of $\chi_{3}$, which mixes 
with the physical Higgs. The imaginary part of $\chi_{3}$ does not mix with the Higgs and is 
therefore stable. However, so long as the $\chi_{3}$ self-coupling $\eta_{3333}$ is large, 
the $\chi_{3}$ scalars will maintain a thermal equilibrium with each other even after they 
have decoupled from thermal equilibrium with other particles. Since the imaginary part of 
$\chi_{3}$ is generally heavier than the real part after the latter mixes with the Higgs, the 
imaginary part of $\chi_3$ will annihilate to the lighter real part once $T < M_{\chi_{3}}$, 
thus ensuring that the entire $\chi_{3}$ density can decay to muons prior to nucleosynthesis. 

\subsection{Sub-dominant $S$ annihilation to Higgs pairs}

           We have so far considered the annihilation $S^{\dagger}S \rightarrow \chi_{3}^{'}
\chi_{3}^{'}$ via the quartic coupling $(\eta_{S})_{33}$, which primarily produces 
$\mu^{+} \mu^{-}$ pairs. However, it is also possible to have $S^{\dagger}S \rightarrow  
H^{\dagger}H$ via the coupling\footnote{We are assuming that the $S$ mass is sufficiently 
large that we can approximately calculate the annihilation rate in the $<H> \rightarrow 0$ 
limit.} $\gamma$.  The branching ratio to Higgs pairs $B_{S^{\dagger}S \rightarrow H^{\dagger}H} 
\approx \gamma^2/(\eta_{S})_{33}^2$ should be small enough that the production of Higgs pairs 
does not result in a large antiproton signal.  This requires that $B_{S^{\dagger}S \rightarrow 
H^{\dagger}H} \lae 0.1$. However, this still allows a significant coupling to Higgs pairs, which 
can contribute a small antiproton component to the cosmic rays from DM annihilation. 
The $S^{\dagger}SH^{\dagger}H$ coupling  also mediates the coupling of $S$ to nucleons, which may 
allow direct detection of DM. These possibilities are distinctive features of Higgs 
portal models which distinguish them from models based on purely leptophilic couplings.

\subsection{Positron Excesses from $S^\dagger S$ Annihilation}

 The annihilation of $S^\dagger S$ pairs will give rise to mostly $\mu^\pm$ pairs which finally 
decay to $e^\pm$ and neutrinos. The electrons and positrons from $S^\dagger S$ annihilation then 
travel under the influence of the galactic magnetic field and therefore the motion of $e^\pm$ is 
expected to be a random walk. As a result a fraction of $e^\pm$ flux will reach the solar system. 

The positron flux in the vicinity of the solar system can be obtained by solving the 
diffusion equation~\cite{se,delahayeetal:2007,cireli&strumia:NPB2008}
\be
\frac{\partial }{\partial t} f_{e^+}(E,\vec{r},t) =
 K_{e^+}(E) \nabla^2 f_{e^+}(E,\vec{r},t) +
\frac{\partial}{\partial t}[b(E) f_{e^+}(E,\vec{r},t)] + Q(E,\vec{r})\;,
\label{diffusion}
\ee
where $f_{e^+}(E,\vec{r},t)$ is the number density of positrons per unit energy $E$, 
$K_{e^+}(E)$ is the diffusion constant, $b(E)$ is the energy-loss rate and $Q(E,\vec{r})$ 
is the positron source term. The positron source term $Q(E,\vec{r})$ from $S^\dagger S$ 
annihilation is given by
\be
Q(E,\vec{r})=n_S^2(\vec{r}) \sigma_S|v_{\rm rel}| \frac{d N_{e^+}}{d E}\;.
\label{source}
\ee
In the above equation the fragmentation function $d N_{e^+}/d E$ represents the
number of positrons with energy $E$ which are produced from the annihilation of
$S^\dagger S$. 

We assume that the positrons are in a steady state, i.e. $\partial f_{e^+}/\partial t=0$. Then from
Eq. (\ref{diffusion}), the positron flux in the vicinity of the solar system can be given in a
semi-analytical form~\cite{se,delahayeetal:2007,cireli&strumia:NPB2008}
\be
\Phi_{e^+} (E,\vec{r}_{\odot})  =  \frac{v_{e^+}}{4\pi b(E)}(n_S)_{\odot}^2
\sigma_S|v_{\rm rel}| \int_E^{M_S} dE' \frac{dN_{e^+}}{dE'}I (\lambda_D(E,E'))\,,
\label{positron_flux}
\ee
where $\lambda_D(E,E')$ is the diffusion length from energy $E'$ to energy $E$ and 
$I(\lambda_D(E,E'))$ is the halo function which is independent of particle physics. An 
analogous solution for electron flux can also be obtained.

\begin{figure}[htbp]
\begin{center}
\epsfig{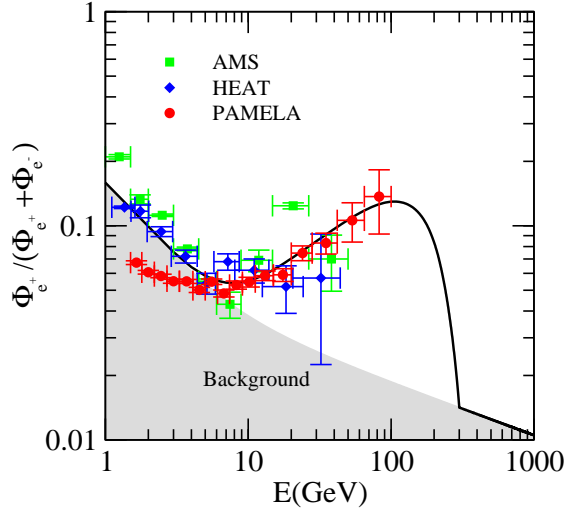}
\caption{Positron fraction from $S^\dagger S \rightarrow 2 \mu^+ \mu^-$ at $M_S=600$ GeV. 
We have used the annihilation cross-section $\langle \sigma|v|\rangle = 4.5 \times 10^{-23} 
{\rm cm}^3/{\rm s}$.}
\label{positron_fitting}
\end{center}
\end{figure}

Positrons in our galaxy are not only produced by $S^\dagger S$ annihilation but also by the scattering of cosmic-ray protons with the interstellar 
medium~\cite{moskalenko&strong:astro1998}. Thus the positrons produced from the latter sources 
can act as background for the positrons produced from the annihilation of $S^\dagger S$. The 
background fluxes~\cite{moskalenko&strong:astro1998} of primary and secondary electrons and 
secondary positrons can be parameterized as~\cite{baltz&edsjo:prd1998}:
\bea
\Phi_{\rm prim,\; e^-}^{\rm bkg} &=& \frac{0.16 \epsilon^{-1.1}}{1+11 \epsilon^{0.9}+3.2\epsilon^{2.15}}
{\rm GeV^{-1} cm^{-2} s^{-1} sr^{-1}}\nonumber\\
\Phi_{\rm sec,\; e^-}^{\rm bkg} &=& \frac{0.70\epsilon^{0.7}}{1+11\epsilon^{1.5}+600 \epsilon^{2.9}
+580\epsilon^{4.2}} {\rm GeV^{-1} cm^{-2} s^{-1} sr^{-1}}\nonumber\\
\Phi_{\rm sec,\; e^+}^{\rm bkg} &=& \frac{4.5 \epsilon^{0.7}}{1+650 \epsilon^{2.3}+1500 \epsilon^{4.2}}
{\rm GeV^{-1} cm^{-2} s^{-1} sr^{-1} }\,,
\eea
where the dimensionless parameter $\epsilon$=E/(1 GeV). The net positron flux in the galactic 
medium is then given by
\be
(\Phi_{e^+})_{\rm Gal}=(\Phi_{e^+})_{\rm bkg} + \Phi_{e^+}(E,\vec{r}_{\odot})
~.\ee
The second term in the above equation is given by Eq. (\ref{positron_flux}), which depends
on various factors: $b(E)$, $\lambda_D(E,E')$, $I (\lambda_D(E,E'))$, $v_{e^+}$, $(n_S)_{\odot}$
and the injection spectrum $dN_{e^+}/dE'$. The energy loss  term $b(E)$ (due to inverse Compton scattering
and synchrotron radiation due to the galactic magnetic field) is determined by the photon
density and the strength of magnetic field. Its value is taken to be
$b(E)=10^{-16} \epsilon^2 {\rm GeV s}^{-1}$~\cite{baltz&edsjo:prd1998}. The number density of
$S$ DM in the solar system is given by
\be
(n_S)_{\odot}=\frac {\rho_\odot}{M_S}
\label{DM_density}
~,\ee
where $\rho_\odot\approx 0.3 {\rm GeV/cm^3}$. In the energy range we are interested in, the
value of $v_{e^+}$ is taken approximately to be $c$, the velocity of light. The values of 
diffusion length $\lambda_D(E,E')$ and the corresponding halo function $I (\lambda_D(E,E'))$
are based on astrophysical assumptions~\cite{se,delahayeetal:2007,cireli&strumia:NPB2008}. 
By considering different heights of the galactic plane and different DM halo profiles the 
results may vary slightly. In the following we take the height of the galactic plane to be $\lsim 4$ kpc, 
which is referred to as the "MED" model~\cite{se,delahayeetal:2007,cireli&strumia:NPB2008}, and we 
have used the NFW DM halo profile~\cite{NFW},
\be
\rho(r)=\rho_{\odot}\left( \frac{r_\odot}{r} \right)\left(\frac{1+ \left( \frac{r_\odot}{r_s} 
\right)
}{1+\left(\frac{r}{r_s} \right)} \right)^2
\label{NFW-profile}
~,\ee
to determine the halo function $I (\lambda_D(E,E'))$, where $r_s\approx 20 {\rm kpc}$ and
$r_{\odot} \approx 8.5 {\rm kpc}$. (We find that our results are not strongly sensitive to the halo profile.) In Figure \ref{positron_fitting}, plotted using 
DARKSUSY~\cite{darksusy}, the positron fraction from $S^\dagger S$ annihilation 
is compared with the data from AMS, HEAT and PAMELA for the case of $M_{S} = 600$ GeV, showing that a good fit is obtained in this case. 

\subsection{Nucleosynthesis, Gamma-Ray and CMB Constraints on Enhanced $S^\dagger S$ Annihilation}

\label{sec:OtherConst}

So far we have considered large annihilation cross sections of the order of $10^{-23}~{\rm
cm}^{3}~{\rm s}^{-1}$ in order to fit the excess of the observed cosmic-ray electron fraction. This value is approximately 
$10^2-10^3$ times larger than the canonical value of the annihilation cross section for thermal relic 
DM ($\simeq 3 \times 10^{-26} {\rm cm}^{3}~{\rm
s}^{-1}$). Therefore we have to check if this value is consistent with
other cosmological and astrophysical constraints, in particular those from nucleosynthesis and due to gamma-rays from the galactic centre (GC) and halo. We will consider $S$ annihilation primarily to $\chi_3^{'}$ pairs, but we will include
the possibility of a small but significant branching ratio to Higgs pairs.

First we shall discuss constraints which come from BBN~\cite{Jedamzik:2004ip,kohri:08,
Hisano:2009rc,Kawasaki:2004yh}. Even after the freeze-out of $S^{\dagger}S$ annihilations, a small 
amount of $S$ pairs continue to annihilate. In our model, the $S^{\dagger}S$ pair dominantly
annihilates into $\mu^{+}\mu^{-}$ pairs with some fraction into $H^{\dagger}H$. 
\begin{figure}[h]
\begin{center}
\epsfig{file=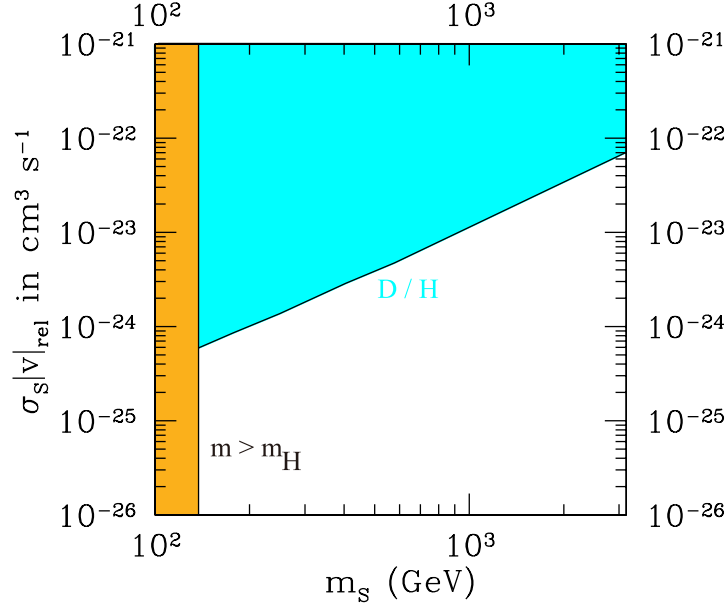, width=0.6\textwidth}
\caption{Upper bounds from BBN on the annihilation cross section of
$S^{\dagger}S$ into a Higgs $H^{\dagger}H$ pair as a
function of the DM mass, where the branching ratio is normalized to $B_{S^{\dagger}S\to
H^{\dagger}H} = 1$. Here we have assumed the mass of Higgs boson is 130
GeV. The name of the light element used for the constraint is written
near each line. The vertical band at the left side indicates the
region which is not kinematically allowed.}
\label{fig:sigma}
\end{center}
\end{figure}

    The photodissociation of D and  $^{4}$He is severely constrained  by observational value of 
$^{3}$He/D. According to \cite{Hisano:2009rc} we have a constraint on the annihilation cross
section into $\mu^{+}\mu^{-}$ pairs,
\begin{equation}
  \label{eq:simgalep}
  \langle \sigma v\rangle <
  \langle \sigma v\rangle_{S^{\dagger}S\to 2\mu^{+}\mu^{-}}^{\rm photo}
  = \frac{1.0 \times 10^{-21}}{B_{S^{\dagger}S\to2\mu^{+}\mu^{-} }}~
  {\rm cm}^3{\rm s}^{-1}
\left(\frac{E_{\rm vis}/M_{\rm S}}{0.7}\right)^{-1}
  \left(\frac{M_{\rm S}}{1\ {\rm TeV}}\right),
\end{equation}
where $E_{\rm vis}/M_{\rm S}$ represents the fraction of the total energy $2 M_{\rm S}$ which 
goes into visible energy $E_{\rm vis}$ i.e. charged particles and photons. $B_{S^{\dagger}S\to 
2\mu^{+}\mu^{-}}$ is the branching ratio for $S^{\dagger}S$ annihilation into 2 $\mu^{+}\mu^{-}$ 
pairs. In case of the muon decay, $E_{\rm vis} \sim 0.7 M_{\rm S}$.  

  In addition, in order to limit the branching ratio to Higgs pairs, we have calculated the constraints 
on the cross section to $H^{\dagger}H$ which follow from photodissociation and hadron
emission. For photodissociation we find \cite{Hisano:2009rc} 
\begin{equation}
  \label{eq:simgalep2}
    \langle \sigma v\rangle <
  \langle \sigma v\rangle_{S^{\dagger}S\to H^{\dagger}H}^{\rm photo}
 = \frac{7.0 \times 10^{-22}}{B_{S^{\dagger}S\to H^{\dagger}H}}~
  {\rm cm}^3{\rm s}^{-1}
\left(\frac{E_{\rm vis}/M_{\rm S}}{1.0}\right)^{-1}
  \left(\frac{M_{\rm S}}{1\ {\rm TeV}}\right),
\end{equation}
where PYTHIA~\cite{Sjostrand:2006za} gives $E_{\rm vis} \sim 1.0 M_{\rm S}$ and and 
$B_{S^{\dagger}S\to H^{\dagger}H}$ is the branching ratio into $H^{\dagger}H$. (In the low 
energy limit this becomes the total branching ratio to $W$, $Z$ and $h$.) In the current 
case $B_{S^{\dagger}S\to H^{\dagger}H} = 1 - B_{S^{\dagger}S\to 2 \mu^{+}\mu^{-}}$. The dominant
upper bound comes from the smaller of $\langle \sigma
v\rangle_{S^{\dagger}S\to2\mu^{+}\mu^{-}}^{\rm photo}$ and $\langle
\sigma v\rangle_{S^{\dagger}S\to H^{\dagger}H}^{\rm photo}$. 
These bounds are generally compatible with the range of values required to account for the cosmic ray excesses, $\langle \sigma v \rangle \sim 3 \times 10^{-24}- 
3 \times 10^{-23} {\rm cm^{3} s^{-1}}$ for boost factors $10^2 - 10^3$. 

The most severe bound on hadron emission comes from the overproduction of deuterium by the destruction 
of $^{4}$He. This process is constrained by observational D/H~\cite{Hisano:2009rc}.
From this we obtain the hadron emission constraint
\begin{equation}
  \label{eq:simgahad}
\langle \sigma
v\rangle <
\langle \sigma
v\rangle_{S^{\dagger}S\to H^{\dagger}H}^{\rm had}
 = \frac{1.3 \times 10^{-23}}{B_{S^{\dagger}S\to H^{\dagger}H}}
~ {\rm cm}^3{\rm s}^{-1}
\left( \frac{N_{n}}{1.0}\right)^{-1}
\left( \frac{M_{\rm S}}{1\ {\rm TeV}}\right)^{1.5},
\end{equation}
with $N_{n}$ the number of emitted neutrons per single annihilation.
In the case of $H^{\dagger}H$ emission, $N_{n}$ is approximately 
1.0, which is obtained using PYTHIA~\cite{Sjostrand:2006za}. This is again consistent with 
the range of $\langle \sigma v \rangle$ required to account for the observed cosmic ray excesses. 
We have plotted the results for annihilation into $H^{\dagger}H$ in Figure 4, with the normalization 
$B_{S^{\dagger}S\to H^{\dagger}H} = 1$.

In summary, for the range of $S$ mass which is compatible with  perturbative couplings, the boost 
factor required to account for the positron and/or electron plus positron excess via annihilation 
to muons is compatible with present BBN constraints. 

We next consider constraints from gamma-rays. A possible gamma-ray signal from the GC due to 
DM annihilation has been extensively studied as it could provide a good method to study the
nature of DM astrophysically. So far the HESS group has reported
that power-law signals were observed from the GC~\cite{Aharonian:2004wa,Aharonian:2006wh} 
for 200 GeV $\lae E_{\gamma} \lae $ 700 GeV. Quite recently the FERMI satellite
group also reported their preliminary result for the signals observed
from the galactic mid-latitude (10$^{\circ}$ < $|b|$ < 20$^{\circ}$)
for 200  MeV $\lae E_{\gamma} \lae$ 10 GeV. 
When we adopt a cuspy profile of the galaxy, such as the NFW
profile, the gamma-ray signal
from muon emission can exceed
the observed signal. However, if we take a milder profile such as the
cored isothermal profile, then for the moment DM annihilation is not
constrained by the current
observations~\cite{Bertone:2008xr,Kawasaki:2009nr,salucci}~\footnote{Note that
the positron and electron plus positron signals will not change even if we used the
cored isothermal profile because local annihilation within 1 kpc dominates the production of electrons and positrons with $\gtrsim$ 10 GeV
energies.}. To clarify the dependence of the DM constraints on the density profile, we need more 
accurate data on the diffuse gamma-ray background, which will be provided by FERMI in the near future.

In addition, there are CMB constraints on the enhanced $S^\dagger S$ annihilation 
cross-section. It has been shown in ref.~\cite{padmanabhan,kanzaki} that energetic particles 
from rapid $S^\dagger S$ annihilation can reionize neutral hydrogen at 
the last scattering surface, leaving an imprint on the CMB. 
The analysis of \cite{padmanabhan} concludes that current data from WMAP5 imposes a 2-$\sigma$ upper bound on the $S^\dagger S$ annihilation cross-section which is given by
\be
\langle \sigma v \rangle_{S^\dagger S \rightarrow \chi_3^\dagger \chi_3} < \frac{3.6 \times 10^{-24} {\rm cm}^3/{\rm s}}{f} \left( \frac{M_S}
{1 {\rm TeV}} \right)\,,
\ee
where $f$ is in the range $0.2-0.3$ for annihilation to 2 $\mu^{+} \mu^{-}$ pairs. 
Thus a boost factor of ${\rm O}(1000)$ is marginally allowed by the current 
data.   

Finally, we briefly comment of the possibility of neutrino signals from the
GC. Detecting such neutrino signals in the future might be useful to
distinguish the Higgs portal DM model from others, since muon neutrinos are produced by the 
decay of the $\mu^{+} \mu^{-}$ pairs coming from DM annihilation and subsequent $\chi_3^{'}$ decay. 
So far Super-K has reported upper bounds on the up-going muon flux coming from neutrinos emitted 
from the GC~\cite{Desai:2004pq}. We can compare the theoretical prediction of the neutrino flux in 
our model with this Super-K upper bound. According to the discussion of Ref.~\cite{Hisano:2008ah}, 
our model is presently allowed since neutrinos are not produced directly but indirectly through the 
decay of the charged leptons and possibly mesons. It is expected that future neutrino experiments 
such as  KM3NeT~\citep{Kappes:2007ci,kappes07} or IceCube DeepCore~\cite{Achterberg:2007qp,Cowen:2008zz} 
will be able to detect the up-going muons induced by the neutrinos emitted from the GC.

\section{A Sommerfeld Enhanced Version of the Model}

   In the previous section we studied the conditions for a successful 
Higgs portal model with non-thermal production of DM. In this section we will 
replace non-thermal production with Sommerfeld enhancement of the annihilation cross-section 
as the source of the boost factor. The main difference between the two models 
is the reduced number of hidden sector scalar fields, since $\chi_{1}$is no longer needed 
as the source of the non-thermal DM density. This will also eliminate the most heavily 
suppressed O($10^{-10}$) couplings, which were necessary to ensure the late decay of $\chi_{1}$. The 
$S$ DM annihilation to $\chi_{3}$ pairs and subsequent $\chi_{3}$ decay to $\mu^{+}\mu^{-}$ 
pairs is unchanged from the non-thermal scenario.

   Since in the non-thermal model there must exist a light scalar $\chi_3$ if we wish to avoid leptophilic couplings, it is natural to ask whether we can eliminate $\chi_1$ and consider instead thermal 
DM with a Sommerfeld enhanced annihilation cross-section, with the attractive force mediated by $\chi_3$-exchange. The correct thermal relic density of $S$ DM is obtained if the $S^\dagger S \chi_{3}^\dagger \chi_{3}$ coupling is in the range 0.1-1 for $M_{S} \sim 0.1-1 \TeV$ \cite{sdm1}.
If we then consider the coupling $(\eta_S)_{23}$ in Eq.(1) and introduce $<\chi_2>$, we obtain the interaction 
\bej{se1}  (\eta_S)_{23}<\chi_2> \chi_3 S^{\dagger}S {\rm \;\;+ \;\; h.c.}  ~.\eej
This interaction can produce the required long-range force between $S$ particles via $\chi_3$ exchange. The condition for a Sommerfeld enhanced annihilation rate is $M_{\chi_3} \lae \alpha M_{S}$, where $\alpha = \lambda^2/4 \pi$ and the effective coupling from $\chi_{3}$ exchange is $\lambda \approx (\eta_S)_{23}<\chi_2>/M_{S}$.  
Therefore
\bej{se3}  M_{\chi_3} \lae 1 \GeV (\eta_{S})_{23}^{2} 
\left(\frac{<\chi_2>}{100 \GeV}\right)^2 \left(\frac{1 \TeV}{M_{S}}\right)     ~.\eej 
Since $M_{\chi_{3}} \sim 200 \MeV$ in our model, this will be satisfied if $(\eta_{S})_{23} \gae 0.4$ when $M_{S} \sim 1 \TeV$. 

   Therefore, in addition to simplifying the model by eliminating $\chi_{1}$, Sommerfeld enhancement permits larger DM masses, $M_{S} \sim 1 \TeV$.  This may be significant in light of recent analyses \cite{na1,na2} of the new FERMI and HESS electron plus positron data, which favour DM particles with 
TeV scale masses which annihilate to muons (with the case of annihilation to 4$\mu$ being favoured by the analysis of \cite{na2}). Since the Higgs portal models generally predict that DM annihilates to two $\mu^{+}\mu^{-}$ pairs via decay of the primary $\chi_3$ pair, a Sommerfeld enhanced version of the Higgs portal model, in contrast with the non-thermal model, could provide an explanation for both the higher energy electron plus positron excess and the lower energy PAMELA positron excess. 

\section{Conclusions}

     We have considered two DM models for cosmic ray excesses which are based on Higgs 
portal-type couplings of a scalar DM sector to the SM, one with non-thermal DM as the 
explanation of the boost factor and the other with thermal DM and Sommerfeld 
enhancement of the annihilation cross-section. 

         In the case of the model with non-thermal production of DM, the DM scalar 
mass must be less than about 600 GeV if the model is to remain perturbative. Therefore if this model 
is correct then the PAMELA positron excess can be explained by DM annihilation but the higher 
energy electron plus positron flux suggested by FERMI and HESS must have a different
explanation. This is likely to be true of most models without Sommerfeld enhancement. 
Non-thermal production of DM is possible via quartic scalar couplings. However, the couplings 
leading to decay of the heavy scalar which produces the DM density must be highly suppressed, 
$\lae 10^{-10}$, in order to ensure that the heavy particle decays well 
after the DM particle freezes-out. 

       A successful model must also account for DM annihilation to primarily leptonic states. If we do not wish to introduce DM which couples preferentially to leptons then the only way to achieve this is kinematically, by ensuring that DM annihilates to unstable final states which are too light to subsequently decay to hadrons. Our model can generate the required decay process via mixing of the $\chi_3$ scalar of the hidden sector with the Higgs, leading to the decay of $\chi_3$ primarily to $\mu^{+} \mu^{-}$ via the muon Yukawa coupling if its mass is in the range $2 m_{\mu} - 2 m_{\pi^{o}}$ ($212-270$ MeV). The small $\chi_3$ mass requires that the quartic scalar couplings of $\chi_3$ to the Higgs and to $\chi_2$ are $\lae 10^{-6}$. The $\mu^{+} \mu^{-}$ final state is essential if we require that the $\chi_3$ density decays prior to nucleosynthesis (which $\chi_3$ would otherwise dominate) 
but does not decay to pions or nucleons, which would produce a potentially dangerous photon or antiproton flux. 
This is a clear prediction of the Higgs portal model, which applies equally to the Sommerfeld enhanced version. 

    We conclude that quartic couplings of a relatively simple scalar DM sector can achieve the required enhancement of the annihilation rate and leptonic final states, but appropriate mixtures of strongly suppressed and unsuppressed quartic couplings and large and small mass terms are required.  In the absence of symmetries or dynamical effects which can explain them, such hierarchies would appear unnatural. It is therefore to be hoped that the pattern of masses and couplings can be understood in terms of the symmetries or dynamics of a complete theory, for which the present model is the low energy effective theory.

   A significant feature of the Higgs portal model, which can distinguish it from those with purely leptophilic annihilation modes, is that there can be a significant coupling of DM to Higgs pairs. This could produce a non-negligible antiproton component in the cosmic rays from DM annihilation if the annihilation process $S^{\dagger}S \rightarrow H^{\dagger}H$ is not too suppressed relative to the dominant process $S^{\dagger}S \rightarrow \chi_3^{\dagger} \chi_3$. The $S^{\dagger}SH^{\dagger}H$ coupling may also allow direct detection of DM.      

     Constraints from BBN are important for the model with non-thermal DM, since the annihilation rate is large at all temperatures. We found that both the muon and Higgs final states are consistent with an annihilation cross section as large as  $10^{-23}~{\rm cm}^{3}~{\rm s}^{-1}$ for $M_{S} \lesssim 600$ GeV. 
The model is also consistent with the gamma-ray signal from the galactic centre and from the diffuse gamma-ray background in the case of a cored isothermal halo profile, but not in the case of a cuspy NFW profile. 

    In the Sommerfeld enhanced version of the model, the low mass $\chi_3$ scalar which accounts for leptonic DM annihilation also mediates the force responsible for the Sommerfeld enhancement. In this case we can reduce the number of additional scalars by one, since $\chi_1$ is no longer needed to produce the DM non-thermally, 
which also eliminates the most highly suppressed couplings. This version of the model can accomodate a TeV scale DM particle, allowing it to explain the electron plus positron excess suggested by FERMI and HESS as well as the positron excess observed by PAMELA. Exactly as in the non-thermal model, DM annihilates to $\chi_3$ pairs which subsequently decay via Higgs mixing  to 2 $\mu^{+}\mu^{-}$ pairs.  This may be significant, as recent analyses suggest that the new FERMI and HESS electron plus positron data favours TeV scale DM particles annihilating to muons \cite{na1,na2}, with annihilation to 4$\mu$ via intermediate decaying scalars being favoured by \cite{na2}.  

     The Higgs portal models considered here should have phenomenological signals due to the coupling of the 
DM sector to the Higgs bilinear and the mixing of the Higgs with the SM singlet $\chi_{2}$ \cite{lhc}. If $S$ or $\chi_{2}$ are light enough then they may be produced via Higgs decay at the LHC. The mass eigenstate Higgs boson is also be expected to have a large singlet component, with consequences for Higgs phenomenology. These features may not be unique to our model, but they would provide indirect support for it. In addition, the muon neutrinos produced by the decay of the $\mu^{+}\mu^{-}$ pair from DM annihilation may be detectable via upward-moving muons at future neutrino experiments.

\section*{Acknowledgement}
The research at Lancaster is supported by
EU grant MRTN-CT-2004-503369 and the Marie Curie Research and
Training Network "UniverseNet" (MRTN-CT-2006-035863).

\end{document}